\documentclass[aip,pof,reprint,graphicx]{revtex4-1}

\usepackage{graphicx}
\usepackage{dcolumn}
\usepackage{bm}
\usepackage{amsmath}
\usepackage[english]{babel}

\draft 

\begin{document}


\title[Numerical simulation of Faraday waves oscillated by two-frequency forcing]
{Numerical simulation of Faraday waves oscillated by two-frequency forcing} 



\author{Kentaro \surname{Takagi}}
\email{kentaro@kyoryu.scphys.kyoto-u.ac.jp}
\affiliation{ Division of physics and astronomy, graduate school
  of science, Kyoto University, Kitashirakawa Oiwaketyo Sakyoku Kyoto
  606-8502 Japan}%
\author{Takeshi \surname{Matsumoto}}%
\email{takeshi@kyoryu.scphys.kyoto-u.ac.jp}
\affiliation{ Division of physics and astronomy, graduate school
  of science, Kyoto University, Kitashirakawa Oiwaketyo Sakyoku Kyoto
  606-8502 Japan}%



\date{\today}

\begin{abstract}
We perform a numerical simulation of Faraday waves
forced with two-frequency oscillations using a level-set method
with Lagrangian-particle corrections (particle level-set method).
After validating the simulation with the linear stability analysis,
we show that square, hexagonal and rhomboidal patterns are reproduced in agreement
with the laboratory experiments [Arbell and Fineberg, Phys.~Rev.~Lett. \textbf{84}, 654 (2000) and
Phys.~Rev.~Lett. \textbf{85}, 756 (2000)].
We also show that the particle level-set's
high degree of conservation of volume is necessary in the simulations.
The numerical results of the rhomboidal states are compared
with weakly nonlinear analysis. Difficulty in simulating other patterns
of the two-frequency forced Faraday waves is discussed.
\end{abstract}

\pacs{}

\maketitle 

\section{Introduction}
Faraday waves~\cite{Faraday1831},
known to exhibit various kinds of crystalline patterns
in simple settings,
have attracted many researchers for about two hundred years.
Faraday waves are the surface waves between two
superposed immiscible fluid layers subjected to a vertical
vibration.
Even recently astounding exotic phenomena continue to be found
in laboratory experiments on Faraday waves.
For example, in Faraday waves with a certain non-Newtonian fluid (shear-thickening fluid),
the behavior of the interface is far beyond what one can imagine from the
interface motion between air and water~\cite{Merkt2004}.
In another surprising experiment, a droplet slightly submerged in a liquid substrate
under a vertical oscillation is found to behave dynamically like a snake~\cite{Pucci2011}.
To physically understand these phenomena, numerical simulations of them,
which may not be possible now,  are expected to play a decisive role.

As a first step to build such numerical methods, we here study numerically
Faraday waves subjected to a two-frequency forcing in a Newtonian fluid.
There are a number of experimental results with this forcing setting
\cite{Edwards1993a,Edwards1994,Muller1993,Kudrolli1998,Arbell1998,Arbell2000,
 Arbell2000a,Arbell2002,Epstein2004,Epstein2008a},
where much richer variations of the selected patterns are found than
in the single-frequency forced cases as listed below.

The study of two-frequency forced Faraday waves starts with the experiments by
Edwards \textit{et al.}\cite{Edwards1993a,Edwards1994} and Muller \cite{Muller1993}.
The two-frequency forcing can be written as
$A_1\cos(m\omega_0 t) + A_2\cos(n\omega_0 t + \phi) $
and characterized by the integers $m$ and $n$.
Edwards \textit{et al.} explored various ratios of the two frequencies such as $m:n = 3:5, 4:5, 6:7, 4:7$
and $8:9$ and mainly investigated the ratio $4:5$.
They observed the quasi pattern, which has a long-range orientational order
but no spatial periodicity.
On the other hand, the experiment by Muller is focused on the driving ratio of $1:2$
and produces a triangular pattern.

In the linear regime of the two-frequency forced case
a bicritical point exists at which
two normal modes with different wavenumber moduli become simultaneously unstable
(for the single-frequency forced case
the bicritical point can be formed by tuning
the frequency of the forcing for shallow layers \cite{Wagner2003}).
The unstable modes then interact with each other nonlinearly.
In the neighborhood of the bicritical point many complex patterns are expected to be
found. A number of experiments around the bicritical point were conducted
by Kudrolli \textit{et al.}\cite{Kudrolli1998},
Arbell \textit{et al.}\cite{Arbell1998,Arbell2000,Arbell2000a,Arbell2002}
and
Epstein \textit{et al.}\cite{Epstein2004,Epstein2008a}.
Kudrolli \textit{et al.} observed patterns that they named superlattice-1 and superlattice-2.
Arbell \textit{et al.} and Epstein \textit{et al.} observed double hexagonal superlattice (DHS),
subharmonic superlattice states (SSS), oscillon, two-mode superlattices (2MS)
and $2k$ rhomboidal states ($2k$R).
Each pattern can be characterized
by the number of excited (discrete) Fourier modes
and
by the nonlinear resonance among them.

To the best of our knowledge, numerical simulation of the two-frequency
forced Faraday waves based on the Navier-Stokes equations solving
the motion of both the top and bottom fluids is reported in this paper
for the first time.
However, such a simulation, not limited to the two-frequency forced case,
requires treatment of the interface with the surface tension force.
We therefore must employ one of the interface-tracking schemes such as
the volume-of-fluid methods, the level-set methods, the front-tracking methods,
(see, e.g., an advanced textbook~\cite{Tryggvason2011}).
In this study, we adopt the level-set method and
investigate whether or not the simulation of two-frequency forced Faraday waves
with the level-set method is consistent with the experimental results.
The reason for adopting the level-set method will be described later.

The first numerical simulation of the single-frequency forced Faraday waves
in three dimensions was performed by P\'erinet \textit{et
  al.}\cite{Perinet2009}, who reproduced
the square and hexagonal patterns in quantitative agreement
with the laboratory experiment by Kityk \textit{et al.}\cite{Kityk2005}.
P\'erinet \textit{et al.}~\cite{Perinet2009} used a front-tracking method.
It is necessary, for example in simulating oscillon or snake-like patterns,
to allow for overturning and topological change of the interface.
We hence believe that other interface-tracking schemes should be explored
and tested for a wider class of the Faraday waves.
Another numerical issue concerns the density difference between the top
and the bottom fluids. In typical laboratory
experiments, these are air and water at room temperature, meaning
three orders of magnitude difference in the densities.
To handle this large difference, it is known that a high quality solver for the
pressure Poisson equation is needed regardless of the choice of
interface-tracking scheme \cite{Tryggvason2011}.
We use a preconditioned BiCGSTAB.

On the theoretical front of the two-frequency forced Faraday waves,
linear stability analysis and weakly nonlinear theory
are available.
Linear analysis was performed by Besson \textit{et al.}\cite{Besson1996},
which is an extension of the single-frequency forced case~\cite{Kumar1994}.
Their results\cite{Besson1996} agree with the experiments quantitatively.
In the weakly nonlinear analysis, whose emphasis is on
the pattern selection of the two-frequency Faraday waves,
Silber \textit{et al.}, Tse \textit{et al.}, Porter \textit{et al.} and Topaz \textit{et al.}
\cite{Silber1999,Silber2000a,Tse2000,Porter2002,Topaz2002,Porter2004,Topaz2004}
formulated an amplitude equation up to third order in amplitude by applying symmetry
based arguments.

By analyzing the structure of the three-wave resonance, they
succeeded in explaining many selected patterns qualitatively.
Quantitative prediction of the pattern can be obtained if the amplitude equation
of the two-frequency forced Faraday waves is derived from the Navier-Stokes equation with
a realistic boundary condition. However this is a formidable task.
A reduced hydrodynamic equation of the two-frequency
Faraday waves was derived by Zhang \textit{et al.}\cite{ZHANG1997a}.
From this reduced equation, the amplitude equations are derived and analyzed
by assuming infinite depth and small viscosity~\cite{ZHANG1997a, Porter2004, Topaz2004}.
Weakly nonlinear analysis based on the Navier-Stokes equations with
infinite depth was carried out by Skeldon \textit{et al.}\cite{Skeldon2007a}.
This approach with realistic amplitude equations is successful in explaining
many patterns observed in the two-frequency forced Faraday waves.
Nevertheless, there are some patterns, such as oscillons\cite{Arbell2000a},
which are not explained so far by the weakly nonlinear analysis.
In the effort to understand these patterns, numerical simulation of the Faraday waves
plays a complementary role.

For this reason, we develop a method of numerical simulation of the two-frequency
forced Faraday waves, which is consistent with the experiments.
Specifically, we here simulate three patterns observed in
the experiments by Arbell \textit{et al.}\cite{Arbell2000a,Arbell2000}.
In particular the rhomboidal pattern does not appear in the single-frequency
forced Faraday waves.
In order to validate the simulations, we compare our results
with the linear stability analysis of two frequency Faraday waves\cite{Besson1996}.
Next, in the nonlinear regime, we reproduce the square pattern and the
hexagonal pattern with the same physical parameters as the respective experiments.
After that, we reproduce and study the rhomboidal state. During the simulations, we
compare two kinds of level-set methods: one is the original implementation \cite{Sussman1994,Sussman1999a}
and the other
is the level-set method with Lagrangian particles (particle level-set method) \cite{Enright2002}.
Finally, we discuss the difficulty of simulating other patterns observed
in the experiments.

The organization of the paper is the following. In Section~\ref{sec:equat-numer-meth},
we describe the fluid dynamical equations of the Faraday waves, the two level-set methods
and numerical discretization of the equations.
The numerical results are presented in Section~\ref{sec:numerical-results}. More specifically,
comparisons of the simulation with the linear analysis and simple patterns such as square
and hexagonal patterns are presented in Section~\ref{sec:comp-with-line} and~\ref{sec:sqhex}.
The simulation of the rhomboidal states is shown in Section~\ref{sec:rhom}.
In Section~\ref{SEC:comp-with-orig}, we compare the original level-set method
and the particle level-set method.
Our summary and discussion are in Section~\ref{sec:concluding-remarks}.

\section{Equations and numerical method}\label{sec:equat-numer-meth}

In this section, we describe our numerical method
for the governing equations and the boundary conditions used for simulating Faraday waves
oscillated by the two-frequency forcing.

\subsection{Navier-Stokes equations}
Faraday waves occur on the interface between an upper and a lower immiscible
fluids. We employ the one-fluid description of the problem. Numerically we simulate
the dynamics in both fluid layers.
The incompressible Navier-Stokes equations are written as
\begin{eqnarray}
  \nabla \cdot \bm{u} &=& 0, \label{eq:con_sh}\\
  \rho D_t\bm{u} &=&- \nabla p + \rho \bm{G} + \nabla \cdot
  \eta ( \nabla \bm{u} +  \nabla \bm{u}^T) + \bm{s}. \label{eq:ns_sh}
\end{eqnarray}
Here, $D_t, p, \bm{u}$ are the material derivative, the pressure and the velocity,
and $\bm{s},~\rho$ and $\eta$ are
the surface force, the density and the viscosity, respectively.
The vector $\bm{G}$ is the gravitational term in the reference frame of the
container,
\begin{eqnarray}
  \bm{G} &=& (- g + A_1 \cos(m \omega_0 t) + A_2 \cos(n \omega_0 t + \theta)) \bm{e}_z\label{vibration}
\end{eqnarray}
where $g, ~ A_1, ~ A_2,~ \omega_0,~\theta,~\bm{e}_z$
are
the gravitational acceleration,
the amplitude of the first periodic forcing,
the amplitude of the second periodic forcing,
the base angular frequency of the periodic forcing,
the phase shift between the two modes,
the unit vector in the vertical $z$-direction.
In this paper, we set the integers $m,~n$ to $2,~3$.
We also use the notations $\omega_1 = m\omega_0 = 2 \omega_0,~\omega_2 = n \omega_0 = 3 \omega_0$.

On the top and bottom boundaries, no-slip boundary conditions are assumed.
For the horizontal direction, we assume periodic boundary conditions.
The interface location $z=\zeta(x,y,t)$ obeys the kinematic boundary condition.
In term of this $\zeta$, the density $\rho$ and $\eta$ are written as:
\begin{eqnarray}
  (\rho, \eta) = \begin{cases}
    (\rho_t, \eta_t) & z > \zeta(x,y,t), \\
    (\rho_b, \eta_b) & z \le \zeta(x,y,t),
  \end{cases}
\end{eqnarray}
where $\rho_t,~\eta_t$ are the density and the viscosity of the top fluid and
$\rho_b,~\eta_b$ are the density and the viscosity of the bottom fluid.

In this sharp interface description, the density and the viscosity change
discontinuously at the dynamically evolving interface.
This situation is a challenge for numerical simulations.
To circumvent this difficulty, various numerical methods have been proposed,
such as the volume-of-fluid methods, the level-set methods and the front-tracking methods,
just to name a few\cite{Tryggvason2011,Andrea2009}.
In this study, we adopt the level-set method.
The reason is as follows.
The level-set method has a high numerical accuracy
of the normal vector and the curvature of the interface,
hence adequate for the gravity-capillary waves.
However, it is well known that the level-set method does not have good mass
conservation properties \cite{Andrea2009}.
A number of improvements have been proposed \cite{Andrea2009,Enright2002,Sussman2000,Sussman1999a}.
Among them, we use the level-set method corrected with Lagrangian particles,
the so-called particle level-set method,
to ensure volume conservation\cite{Enright2002}.
This conservation problem is discussed in detail in Section~\ref{SEC:comp-with-orig}.

In the following section~\ref{sec:level-set-method}, we describe the level-set method without the particles, here we call
the original level-set method, and  the particle level-set method
and their numerical discretizations.
The description of the discretization of the Navier-Stokes equations
follows later.

\subsection{level-set method}\label{sec:level-set-method}

\subsubsection{level-set function}
We use the level-set approach \cite{Sussman1994} to describe the interface motion.
Here the level-set function $\phi(\bm{x}, t)$, the signed distance from the interface,
indicates the interface. We define $\phi > 0$ in the top fluid and $\phi < 0$ in
the bottom fluid.
The level-set function obeys the following equation
\begin{eqnarray}
  \partial_t \phi + (\bm{u} \cdot \nabla) \phi = 0, \label{eq:ls_main}
\end{eqnarray}
which is discretized with the 5th-order WENO scheme\cite{Jiang2000}
and integrated
in time through the 3rd-order TVD Rung-Kutta method\cite{Jiang2000}.

\subsubsection{Reinitialization of level-set function}\label{redistance}
It is known that the analytic integration of Eq.~(\ref{eq:ls_main}) does not
ensure that $\phi({\bm x}, t)$ is the signed distance function from the interface.
By definition, being the signed distance function requires $|\nabla \phi| = 1$. However
this unit gradient condition is not satisfied since
the Lagrange derivative of $|\nabla \phi|^2$ is not zero but
$D_t |\nabla \phi|^2  =  - 2 (\nabla \phi)(\nabla  {\bm u})(\nabla \phi)$.
To enforce the condition (in practice, we do so just around the interface),
the distance function is re-initialized at each time step
from the following initial value problem with the virtual time $\tau$
\cite{Sussman1999a}
\begin{eqnarray}
 && \frac{\partial d}{\partial \tau} = \mathrm{sgn}(\phi)(1-|\nabla d |) +
  \lambda f(\phi), \label{eq:ls_reinit} \\
 && d(x,y,z,\tau=0) = \phi(x,y,z). \notag
\end{eqnarray}
Although we call $\tau$ virtual time, its dimension is length.
Ideally, the function $d(\bm{x}, \tau)$ as $\tau \to \infty$
gives the corrected signed distance function for all the computational domain.
Here, we set $\phi(\bm{x}, t) = d(\bm{x}, \tau = \tau_l)$ for
some value $\tau_l$. This $\tau_l$ corresponds to the largest distance from the interface
to which we demand $\phi$ be the signed distance.
In this paper, we use $\tau_l = \epsilon$, where
$\epsilon$ is the half width of the diffuse interface
and set to $2 \Delta z$, where $\Delta z$ is the grid spacing in the vertical $z$-direction.
The functions $\lambda(\bm{x}), f(\phi)$ in Eq.~(\ref{eq:ls_reinit}) are given as
\begin{eqnarray}
  \lambda(\bm{x}) &=& - \frac{\int_{\Omega(\bm{x})} H'(\phi) L(\phi,d)
    d\bm{x}}{\int_{\Omega(\bm{x})} H'(\phi) f(\phi) d\bm{x}}, \\
  f(\phi) &=& H'(\phi)|\nabla \phi|,\\
  H'(\phi) &=& \frac{d H}{d\phi},\label{eq:derivative-heavi-side}\\
  H(\phi) &=&
   \begin{cases}
    0  &\text{if }~ \phi < - \epsilon,\\
    \frac{1}{2}\{1 + \frac{\phi}{\epsilon} +
    \frac{1}{\pi}\sin(\frac{\pi\phi}{\epsilon})\} &\text{if}
    -\epsilon \le \phi \le \epsilon, \\
    1 &\text{if}~ \phi > \epsilon,
   \end{cases}\label{eq:2}
\end{eqnarray}
where $\Omega(\bm{x})$ is a small region centered at the point $\bm{x}$,
$L(\phi, d) = \mathrm{sgn}(\phi)(1 - |\nabla d|)$,
$\epsilon = 2\Delta z$ is the prescribed interface width,
$H(\phi)$ is the smoothed Heaviside function and
$H'(\phi)$ is the smoothed delta function.

Numerically, the reinitialization is done in the following way.
Firstly, we ignore the term $\lambda(\bm{x}) f(\phi)$
in the Eq.~(\ref{eq:ls_reinit}) and solve
\begin{eqnarray}
  \frac{\partial d}{\partial \tau} = \mathrm{sgn}(\phi)(1-|\nabla d |),
\end{eqnarray}
where the discretization in space
is the same as that of Eq.~(\ref{eq:ls_main}).
The integration in the virtual time is discretized as follows,
\begin{eqnarray}
  d' &=& d^n + \Delta \tau L(\phi,d^n)\\
  d^* &=& \frac{1}{4}(3d^n + d' + \Delta\tau L(\phi,d'))\\
  d'^{n+1} &=& \frac{1}{3}(d^n + 2(d^* + \Delta\tau L(\phi,d^*))),
\end{eqnarray}
where $\Delta \tau = 0.5 \min(\Delta x, \Delta y, \Delta z)$.

Secondly, we calculate $\lambda(\bm{x})$
according to the following equation
\begin{eqnarray}
  \lambda_{ijk} =
   \frac{- \int_{\Omega_{ijk}}
    H'(\phi)
    \frac{d'^{n+1}-\phi}{\Delta\tau}
    d\bm{x}}
  {\int_{\Omega_{ijk}}H'(\phi) f(\phi) d\bm{x}}.
\end{eqnarray}
Here $\lambda_{ijk}$ denotes $\lambda(\bm{x}_{ijk})$ on the
grid point ${\bm x}_{ijk}$ specified by the index $(i, j, k)$.
The integral range $\Omega_{ijk}$ describes
the cell region
associated with the grid point.
For the three dimensional case,
by following the two-dimensional version \cite{Takahira2004},
we discretize the integral of some function $g({\bm x})$ in the cell as
\begin{eqnarray}
  \int_{\Omega_{ijk}} g(\bm{x}) ~d\bm{x} &=& \frac{\Delta x \Delta y \Delta
    z}{1512} [9(g_{i+1,j+1,k} + g_{i+1,j-1,k} \nonumber\\
    && + g_{i-1,j+1,k} + g_{i-1,j-1,k} + g_{i,j+1,k+1}  \nonumber\\
    && + g_{i,j+1,k-1} + g_{i,j-1,k+1} + g_{i,j-1,k-1} \nonumber \\
    && + g_{i+1,j,k+1} + g_{i+1,j,k-1} + g_{i-1,j,k+1} \nonumber\\
    && + g_{i-1,j,k-1}) + 88 (g_{i+1,j,k} + g_{i-1,j,k} \nonumber\\
    && + g_{i,j+1,k} + g_{i,j-1,k} + g_{i,j,k+1} \nonumber\\
    && +  g_{i,j,k-1} ) + 876 g_{i,j,k}] ,
\end{eqnarray}
where $\Delta x, \Delta y, \Delta z$ are the grid spacings along
the $x, y, z$ directions.

Finally, $d^{n+1}$ is calculated by
\begin{eqnarray}
  d^{n+1} &=& d'^{n+1} + \Delta\tau \lambda_{ijk} H'(\phi) |\nabla\phi|.
\end{eqnarray}
In practice, we take the total number of the virtual time steps as
$\tau_l /\Delta \tau \simeq 4$.

\subsection{Particle level-set method}

In order to improve the volume conservation of the level-set method,
it has been proposed
to utilize Lagrangian information to correct the level-set function
by adding marker particles near the interface.
Our procedure of the particle level-set method
is basically the same as that of Enright \textit{et al.}\cite{Enright2002}.
The differences are in the error correction and the reseeding strategy.

\subsubsection{Initialization of particles}
The marker particles are spread in the neighborhood of the interface, in which
$|\phi|< 3 \text{max}(\Delta x, \Delta y,\Delta z)$ is satisfied.
The number of particles in each cell is set to 64.
A marker particle has sign $s_p = 1$ or $-1$ and the radius $r_p$.
There are a number of strategies for setting the sign and radius. One simple strategy
is to set the sign to that of the level-set function at the particle
and the radius to the absolute value of the level-set function.
However, we follow the more sophisticated strategy proposed by Enright \textit{et
  al.} to improve numerical results.

The strategy is as follows.
Initially, the particle's sign is set randomly.
In order to have the same sign between the particle and the level-set function,
the particles at $\bm{x}_p,~ (p = 1, 2, 3, \ldots)$ are iteratively moved by
the following recurrence relation
\begin{eqnarray}
  \bm{x}_p^{n+1} &=& \bm{x}_p^n + 2^{-n} (\phi_{goal} -\phi(\bm{x}^n))
  \bm{N}(\bm{x}_p^n) \label{eq:ls_particle_init},
\end{eqnarray}
where ${\bm N}(\bm{x}_p^n) = \frac{\nabla \phi(\bm{x}_p^n)}{\lvert \nabla \phi(\bm{x}_p^n) \lvert}$ is the normal vector.
Here $\phi_{goal}$ is set as follows.
The sign, $\mathrm{sgn}(\phi_p)$,
is set to have the same as that of the particle $s_p$.
In addition, the absolute value is chosen to be a uniformly distributed random variable
in the range $b_{\min}< |\phi_{goal}| < b_{\max}$.
In this study, $b_{\min}$ is set to $0.1 \min(\Delta x, \Delta y, \Delta z)$ and
$b_{\max}$ is set to $3 \max(\Delta x, \Delta y, \Delta z)$.
Each particle is moved repeatedly by Eq.~(\ref{eq:ls_particle_init})
until it satisfies the condition $b_{\min}< s_p \phi(\bm{x}_p) <b_{\max}$.
Finally, each particle radius is set according to
\begin{eqnarray}
  r_p=\begin{cases}
    r_u  &  s_p \phi(\bm{x}_p) > r_u,\\
    s_p\phi(\bm{x}_p)  &  r_l<s_p \phi(\bm{x}_p) < r_u,\\
    r_l  &  s_p \phi(\bm{x}_p) < r_l,\\
  \end{cases}
\end{eqnarray}
where $r_{l}$ and $r_{u}$ are lower and upper limits of particle radius
to prevent the creation of particles which are too small or too large.
We use $r_l = 0.1 \min(\Delta x, \Delta y, \Delta z)$ and
$r_u = 5 r_{l}$.
The particle radius $r_p$ is used to correct the level-set function later.
After this procedure, the positive particles at position ${\bm x}_p$
are in the $\phi({\bm x}_p) > 0$ side (the top fluid)
and the negative particles are in the $\phi({\bm x}_p) < 0$ side (the bottom
fluid).
The envelope formed by the circles of the same-sign particles coincides with the
interface $\phi=0$.

\subsubsection{Advection of particles}
Each particle at position ${\bm x_p}(t)$ is advected by
\begin{eqnarray}
  \frac{d \bm{x}_p(t)}{dt} = \bm{u}(\bm{x}_p(t), t).\label{eq:particle_advect}
\end{eqnarray}
The velocity at the particle position $\bm{u}(\bm{x}_p(t), t)$ is calculated
with trilinear interpolation from the velocity vectors on the nearby
cell faces.
The 3rd order TVD Runge-Kutta method is used
to integrate Eq.~(\ref{eq:particle_advect}) in time.

\subsubsection{Error correction of level-set function}\label{error_correction}
As a result of the advection Eq.~(\ref{eq:particle_advect}), some particles move across the interface
$\phi = 0$.
Such escaped particles  are used to
correct the level-set function $\phi$ in the following manner.
First, particles placed on the wrong side ($\phi(\bm{x_p}) \times s_p < 0$)
are considered to have escaped.
Second, we introduce the signed distance function
between the escaped particle and a point $\bm{x}$,
which is calculated with the particle radius $r_p$ as
\begin{eqnarray}
  \phi_p(\bm{x}) = s_p(r_p - |\bm{x}-\bm{x}_p|).
\end{eqnarray}
This signed distance is positive ($\phi_p(\bm{x}) > 0$)
if the point $\bm{x}$ is within the positive ball ($s_p=1$) of radius $r_p$
centered on $\bm{x}_p$.
The distance function corrected by the escaped positive (negative) particle
$\phi_+(\bm{x}) ~(\phi_-(\bm{x}))$ is calculated from
\begin{eqnarray}
  \phi_+(\bm{x}) = \underset{\forall p \in E^+}{\text{max}} (\phi_p,\phi(\bm{x})), \\
  \phi_-(\bm{x}) = \underset{\forall p \in E^-}{\text{min}} (\phi_p,\phi(\bm{x})).
\end{eqnarray}
Here $E^{+}$ and $E^{-}$ denote the sets of the escaped positive and negative particles.
Finally, the level-set function $\phi(\bm{x})$ is corrected as
\begin{eqnarray}
  \phi(\bm{x}) = \begin{cases}
    \phi_+(\bm{x}) & \text{if} ~ |\phi_+(\bm{x})| \le |\phi_-(\bm{x})|, \\
    \phi_-(\bm{x}) & \text{if} ~ |\phi_-(\bm{x})| < |\phi_+(\bm{x})|.
  \end{cases}
\end{eqnarray}
Ideally, after this correction of the distance function, all the particles tagged as escaped
have the same sign as the corrected distance function.
However, with our implementation of the correction in the preliminary calculations,
we find that some particles do not have the same sign of the corrected
distance function. If we use such particles with the wrong sign in the next correction process,
the interface becomes nearly singular, which we consider a numerical artifact. Therefore we ignore
such escaped particles in the later correction processes.
The point differs from the usual procedure of the particle level-set method\cite{Enright2002}.

\subsubsection{Reseeding of particles}
Generally, as a result of advection of the particles by a flow,
some regions lack sufficient particles to correct the level-set function.
We reseed the particles where needed.
Specifically, in the cells near the interface ($|\phi(\bm{x})| \le b_{\max}$),
we keep the number of particles in a cell to 64
by adding particles for cells which particles exit
or deleting particles for cells which particles enter.
In our simulation, the reseeding procedure is executed with
the following two strategies.
The first strategy is that the reseeding is done after 40 time steps from the previous
reseeding.
The second strategy is that the reseeding is done when the surface
area of the interface increases by 30\% after the previous reseeding time.
The surface area is calculated from
\begin{eqnarray}
  A&=&\int \delta(\phi) |\nabla \phi| d\bm{x}, \label{eq:surface}\\
  \delta(\phi) &=&
  \begin{cases}
    0 & |\phi| > \epsilon,\\
    \frac{1}{2 \epsilon} \left(1 + \cos(\frac{\pi
        \phi}{\epsilon})\right) & |\phi| \le \epsilon,
  \end{cases}
\end{eqnarray}
where the smoothed delta function $\delta(\phi)$ is the same as $H'(\phi)$ of
Eq.~(\ref{eq:derivative-heavi-side}).

In summary, the one-step update of the level-set function with particles
is carried out by the following steps\cite{Enright2002}.
\begin{enumerate}
\item The level-set function is advected by Eq.~(\ref{eq:ls_main}).
\item The particles are advected by Eq.~(\ref{eq:particle_advect}).
\item The error of the level-set function is corrected by the procedure described in the Section~\ref{error_correction}.
\item The level-set function is reinitialized as described in Section~\ref{redistance}.
\item The error of the level-set function is once more corrected by the particles as described in Section~\ref{error_correction}.
\end{enumerate}
For the original level-set method without particles, the second, third and fifth
processes are omitted.

\subsection{Discretization of Navier-Stokes Equations}

We use the following temporal discretization of the incompressible
Navier-Stokes Eqs.~(\ref{eq:con_sh}) and (\ref{eq:ns_sh})
with the projection method and with adaptive time stepping
\begin{eqnarray}
  \bm{u}^* =
   &&\bm{u}^n
   + \Delta t^n
     \bigg[\nonumber \\
  && - \left(
    1
    + \frac{\Delta t^{n-1}}{\Delta t^n}\right)
          {\bm A}^n
        + \frac{\Delta t^{n-1}}{\Delta t^n}
      {\bm A}^{n-1} \nonumber \\
  && + \left(
          1
         + \frac{\Delta t^{n-1}}{\Delta t^n}
       \right)
       {\bm D}_e^n
     - \frac{\Delta t^{n-1}}{\Delta t^n}
       {\bm D}_e^{n-1} \nonumber \\
  && +  \frac{1}{2}
         \left(
        {\bm D}_i^*
      + {\bm D}_i^n
     \right)
     + {\bm G}^{n}
     + {\bm S}^{n+1}
     \bigg],\label{update_velocity}\\
      \bm{u}^{n+1}
       =
   &&    \bm{u}^*
       + \Delta t^n
          \frac{1}{\rho^n}\nabla p^{n+1}.\label{update_pressure}
\end{eqnarray}
Here the superscript $n$ denotes the value at the $n$-th time step
$t^n = \sum_{i = 1}^{n} \Delta t^{i}$ in which $\Delta t^i$ is the time step size for the $i$-th
step. We discuss later how to determine them.
${\bm A}^n$ is the advective term, ${\bm D}_i^n$ and ${\bm D}_i^*$ are
the viscous terms involving the same component
as on the left-hand-side and at the intermediate step, ${\bm D}_e^n$ is
the viscous term involving the other components,
${\bm G}^n$ is gravity and ${\bm S}^n$ is the surface force term.
Here, by $^*$ we denote the intermediate step.
As in the standard way of the projection method,
the pressure term $p^{n+1}$ is
calculated from the divergence free condition.
Equation~(\ref{update_pressure}) acted upon by $\nabla \cdot$ becomes
\begin{eqnarray}
  - \frac{\nabla \bm{u}^*}{\Delta t^n} =\nabla\left( \frac{1}{\rho^n} \nabla p^{n+1}\right).
\label{ppoi}
\end{eqnarray}
The advection term $\bm{A}^n$ is described by
\begin{eqnarray}
  \bm{A}^n = \bm{u}^n \cdot \nabla \bm{u}^n.
\end{eqnarray}
The $x$-component of the viscous term $\bm{D}_i^n$ to be treated implicitly is
\begin{eqnarray}
  D_{ix}^n = \frac{1}{\rho^n} \left[\nabla \left(\eta^n \nabla u_x^n \right) +
    \frac{\partial}{\partial x}\left(\eta^n \frac{\partial u_x^n}{\partial
        x}\right)\right].
\end{eqnarray}
The viscous term $D_{ix}^*$ is defined similarly but with the intermediate velocity ${\bm u}^*$.
The $x$-component of the viscous term $\bm{D}_e^n$ to be treated explicitly is
\begin{eqnarray}
  D_{ex}^n = \frac{1}{\rho^n}\left[\frac{\partial }{\partial y} \left(\eta^n
      \frac{\partial u_y^n}{\partial x}\right)
  + \frac{\partial }{\partial z} \left(\eta^n \frac{\partial u_z^n}{\partial
      x}\right)\right].
\end{eqnarray}
Other components of the viscous terms are described in the same manner.
The gravitational term $\bm{G}^n$ is
\begin{eqnarray}
  \bm{G}^n
   &=&
   (- g
    + A_1 \cos(\omega_1 t^n)
    + A_2 \cos(\omega_2 t^n +  \theta)
    )
   \bm{e}_z.
\end{eqnarray}
The surface force term $\bm{S}^n$ is
\begin{eqnarray}
  \bm{S}^n = \sigma \kappa^n \bm{n}^n,\quad
  \bm{n}^n = \frac{\nabla \phi^n}{|\nabla \phi^n|}, \quad
  \kappa^n = \nabla \cdot \bm{n}^n.
\end{eqnarray}
Regarding the spacial discretizations,
the advection term, ${\bm A}^n$, is discretized
with the 2nd-order ENO scheme.
The other derivative terms in
${\bm D}_i^*, {\bm D}^n_i, {\bm D}_e^n, {\bm n}^n, \kappa^n$
are discretized with the 2nd-order central difference scheme.

Concerning the boundary conditions,
we assume periodic boundary conditions in the
horizontal directions ($x$ and $y$ directions).
For the vertical $z$ direction, we assume the non-slip condition at $z = 0, L_z$
\begin{eqnarray}
  \bm{u}|_{z=0,\,L_z} = {\bm 0}.
\end{eqnarray}
The boundary condition for the pressure in solving the Poisson equation (\ref{ppoi})
is
\begin{eqnarray}
  \left.\frac{\partial p}{\partial z}\right|_{z=0,\,L_z} = 0.
\end{eqnarray}
For calculating ${\bm u}^*$ in Eq.~(\ref{update_velocity}), the localized ILU preconditioned BiCGSTAB method\cite{Saad2003}
is adopted and the Poisson equation of the pressure is solved by the
multigrid preconditioned BiCGSTAB method\cite{Saad2003}.

Finally, we describe how we determine the variable time step size $\Delta t^i$
which is determined by
\begin{eqnarray}
 \Delta t^i = c \min_{{\bm x}}(\Delta t_S,\, \Delta t_f^i(\bm x),\, \Delta t_{\eta}(\bm x),\,\Delta t_{cfl}^i(\bm x)),
\end{eqnarray}
where we set the safety constant $c = 0.40$.
Here $\Delta t_S$, $\Delta t_f$ and $\Delta t_{\eta}$ are
the time scales of surface force, the vertical vibration and the viscosity, respectively.
The time scale $\Delta t_{cfl}$ concerns the CFL condition.
These reference time scales are defined as
\begin{eqnarray}
&&   \Delta t_S = \sqrt{\frac{(\rho_t + \rho_b) dh^3 }{4\pi \sigma}}, \quad
    (dh = \min(\Delta x, \Delta y, \Delta z)), \\
&&  \Delta t_f^i(\bm x) = \sqrt{\frac{dh}{\bm{G}^i \cdot \bm{e}_z}}, \quad
    \Delta t_{\eta}(\bm x) = \frac{\rho}{\eta}\frac{1}{\frac{1}{\Delta x^2} +\frac{1}{\Delta y^2} + \frac{1}{\Delta z^2}}, \\
&&  \Delta t_{cfl}^i(\bm x) = \frac{1} {\frac{u_x^i}{\Delta x} +\frac{u_y^i}{\Delta y} + \frac{u_z^i}{\Delta z}}.
\end{eqnarray}
Typically $\Delta t_S$ is the smallest in our all simulations.

\section{Numerical Results}\label{sec:numerical-results}

Our goal in this paper is numerical simulation of the $2k$ rhomboidal states
observed in the laboratory experiments by Arbell \textit{et al.}\cite{Arbell2000,Arbell2000a}.
To the best of our knowledge, the rhomboidal states
have not previously been obtained in numerical simulations of the Navier-Stokes equations.
In particular, we use the same bulk fluid parameters as the experiments.
The only difference is the geometry of the system,
i.e., the domain size and the boundary conditions.
In the simulations we apply
periodic boundary conditions in the horizontal directions,
with which we can reduce numerical cost by not simulating
many repeated patterns in the computational domain.
While the experiments are conducted in an open container,
we assume the presence of a rigid wall above the top fluid, on which
the no-slip boundary condition is applied.
This setting is numerically easier than simulating the top-fluid motion in
a semi-infinite domain.
We assume that the top-fluid height is four times larger than
the bottom-fluid depth.

Before presenting the simulation, we first describe the validation of our
fully nonlinear simulation with the above mentioned
geometry by comparing with the linear stability analysis of two-frequency forced
Faraday waves \cite{Besson1996}. This validation process is the same as
P\'erinet \textit{et al.} \cite{Perinet2009}.
The second test then is to reproduce the square and hexagonal patterns
observed in the same two-frequency forced experiments \cite{Arbell2000,Arbell2000a}.
A direct numerical simulation of the square and hexagon patterns for
the single-frequency forced Faraday waves is performed by
P\'erinet \textit{et al.} \cite{Perinet2012a,Perinet2009}. The result on the
rhomboidal states is presented after the validations.

\subsection{Comparison with linear stability analysis}
\label{sec:comp-with-line}
\begin{figure}[h]
  \centering
  \includegraphics[width=8.5cm]{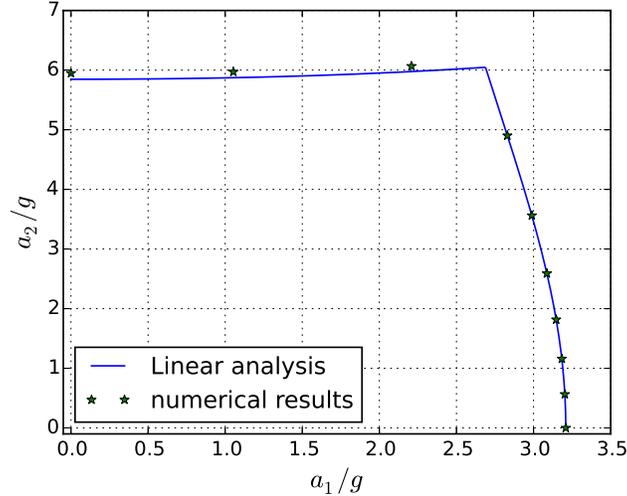}
  \caption{Comparison of the critical vibration amplitudes obtained with the numerical simulation
  to those calculated with the linear stability analysis.}\label{fig:linear_analysis}
\end{figure}

We now compare the critical amplitudes of the oscillations calculated with
the fully nonlinear numerical simulation with those calculated with the
linear stability analysis.
We write the two-frequency forcing as
$a(\cos(\chi)\cos(\omega_1 t) + \sin(\chi)\cos(\omega_2 t+\theta))$.
In the linear analysis,
once we fix the physical parameters as shown in Table~\ref{tab:linear-stability-parameter}
and the mixing angle $\chi$, then the critical value of $a$, denoted as $a_c$,
and the associated critical wave number can be calculated \cite{Kumar1994,Besson1996}.
Here we assume that either harmonic frequency ($\omega_1, \omega_2$) or sub-harmonic frequency $(\omega_1/2, \omega_2/2)$
gives the lowest critical value.
The critical amplitudes $a_{1c} = a_c \cos(\chi)$ and $a_{2c} = a_c \sin(\chi)$ for
the mixing angle $\chi$ from $0^{\circ}$ to $90^{\circ}$ are shown as the solid line in Fig.~\ref{fig:linear_analysis}.

Meanwhile, with the particle level-set simulation, we determine the critical amplitudes
by adding small perturbations to basic modes for ten different values of the mixing angle,
$\chi = 0^{\circ}, 10^{\circ}, 20^{\circ}, \dots, 90^{\circ}$.
The results are denoted as points in Fig.~\ref{fig:linear_analysis}.
The way to estimate the critical amplitudes $a_{1c}$ and $a_{2c}$ in the nonlinear
simulation is as follows:
(i) the perturbation is added to the normal mode whose wavenumber
is set to the critical wave number ($k_c$) calculated from the linear analysis.
More precisely, the perturbation, $\xi \sin(k_c x)$, is added to the flat interface,
where the perturbation amplitude $\xi$ is set to $2 \times 10^{-2}L_z$.
(ii) the interface height $\zeta(x,y,t)$ at the center of
the calculation domain is monitored throughout the simulation for given
$a$ and $\chi$.
The interface height $z = \zeta(x,y,t)$ is calculated from
the zero points of the level-set function $\phi(x,y,z,t)=0$.
We perform such simulations by changing $a$ and estimate the critical value $a_c$.
More precisely, we take the absolute relative difference between
the two peak interface heights at $t = 4.44 T_v$ and $6.44 T_v$ where $T_v = 2
\pi / \omega_0 = 2 \pi / (\omega_1 /2) = 2 \pi /(\omega_2 / 3)$.
Then, $2T_v$ is the minimal period of the two waves with the subharmonic frequencies
$\omega_1 /2$ and $\omega_2 / 2$. If the difference is smaller
than $10^{-3}$, this $a$ is regarded as the critical amplitude of the level-set
numerical simulation.

As shown in Fig.~\ref{fig:linear_analysis},
the critical amplitudes calculated with the level-set simulation tend to be
greater than those calculated with the linear analysis in the $a_2$ dominant region,
namely $a_1/g < 2.6$.
The absolute relative error between the linear analysis (line) and the simulation (point) in the
region is about $0.02$. The agreement between the two results is hence obtained with
two-digit accuracy.

\begin{table}
  \caption{\label{tab:linear-stability-parameter} Parameter values for the linear
    stability analysis.
    These parameters (except for $L_z$ and $\theta$) are identical to the
    experiment by Arbell \textit{et al.}\cite{Arbell2000a}
 }
  \begin{ruledtabular}
    \begin{tabular}{ccl}
      $\rho_t$                & $1.293 $                & ${\rm [kg \, m^{-3}]}$       \\
      $\rho_b$                & $9.500\times 10^{2} $   & ${\rm [kg \,m^{-3}]}$       \\
      $\eta_t$                & $1.822 \times 10^{-5} $ & ${\rm [kg \,m^{-1} \,s^{-1}]}$ \\
      $\eta_b$                & $2.185 \times 10^{-2} $ & ${\rm [kg \,m^{-1} \,s^{-1}]}$ \\
      $\omega_1 = 2 \omega_0$ & $3.770 \times 10^{2} $  & ${\rm [s^{-1}]}$         \\
      $\omega_2 = 3 \omega_0$ & $5.655 \times 10^{2} $  & ${\rm [s^{-1}]}$         \\
      $\theta$                & $0 $                    & ${\rm [rad]}$            \\
      $\sigma $               & $2.150 \times 10^{-2} $ & ${\rm [kg \,m^{-1}]}$           \\
      $g$                     & $9.807 $                & ${\rm [m \,s^{-2}]}$          \\
      $L_z$                   & $1.00 \times 10^{-2} $  & ${\rm [m]}$              \\
      bottom-fluid depth      & $2.00 \times 10^{-3} $  & ${\rm [m]}$
    \end{tabular}
  \end{ruledtabular}
\end{table}

\subsection{Square and hexagonal patterns}
\label{sec:sqhex}

Having validated the simulation of the two-frequency forced Faraday waves
in the linear regime,
we now move to two nonlinear cases: the square and hexagonal patterns.
Note that the two patterns are also observed in the single-frequency Faraday
waves.

First we reproduce the square pattern observed in the experiments by Arbell \textit{et
al}.\cite{Arbell2000,Arbell2000a}.
The physical parameters are shown in Table~\ref{tab:square-parameter}.
We select the amplitudes of the forcing $A_1$ and $A_2$
according to the following reasons:
(i) we aim to conduct the simulation in the weakly nonlinear regime;
(ii) we aim to set the values to be neither close to nor far from the bicritical point.
The selected values of $A_1$ and $A_2$ in Table~\ref{tab:square-parameter}
are of course in the square-pattern domain of the phase diagram
obtained experimentally \cite{Arbell2000a}.
However the geometry of the simulation is different.
We set the lateral dimensions of the computational domain so that it includes
one square $2\pi / k_1 = \lambda_1$, where $k_1$ is the critical wave number, which is
found to be $1.436 \times 10^3~[{\rm m^{-1}}]$ from the linear stability analysis described
in the previous section.
In other words, we set the computational domain to a square box with $L_x = L_y
= \lambda_1$.
This setting is the minimal computational domain
which supports the periodic square pattern. The number of grid points used
in each horizontal direction is denoted by $N_x = N_y$.
The square pattern consists of the four discrete
Fourier modes shown as black dots in Fig.~\ref{fig:nonlinear-resonance}(a).
These modes, called resonant modes, are on the circle of radius $k_1$.
The amplitude of the resonant wavevectors can be calculated from the linear stability
analysis. The direction of those can be estimated from
the experimental data.
The experimental information of the direction is trivial in the case of the square pattern.
However, the information becomes crucial in the case of more complex patterns as
we will see later.
The resultant grid on the Fourier space is shown in Fig.~\ref{fig:nonlinear-resonance}(a).

We start the simulation with zero velocity everywhere and the perturbed flat interface.
The perturbation of the interface is given in terms of the Fourier modes
$\hat{\zeta}({\bm k}, t=0)$ for the wavenumber range $0<|k_x|< 0.5 \max(k_x) =
\pi N_x/(2 L_x) $ and
$0<|k_y|<0.5\max(k_y)=\pi N_y /(2 L_y)$.
The real and imaginary parts of $\hat{\zeta}({\bm
    k})$ in the range are set by independently and identically distributed random
variables with a uniform distribution between $-1/2$ and $1/2$. The zero
Fourier mode is set $\hat{\zeta}({\bm k}={\bm 0}) = 0.2L_z$.
We lastly
transform $\hat{\zeta}({\bm k})$ in the physical space and multiply the perturbation by
an arbitrary factor so that $\max(|\zeta(\bm{x}) - 0.2L_z|) = 5.0 \times 10^{-2}L_z$.

We use here both the original level-set method and the particle level-set method
for comparison.

\begin{table}
  \caption{\label{tab:square-parameter} Parameter values for the square
    pattern. The number of the grid points for each direction are $N_x, N_y, N_z$.
    The other parameters are the same as the Table~\ref{tab:linear-stability-parameter}.
  }
  \begin{ruledtabular}
    \begin{tabular}{ccl}
      $L_x$                       & $4.373\times 10^{-3}$    & ${\rm [m]}$\\
      $L_y$                       & $4.373\times 10^{-3}$    & ${\rm [m]}$\\
      $A_1$                       & $2.000 \times 10^{1}$    & ${\rm [m\,s^{-2}]}$\\
      $A_2$                       & $6.000 \times 10^{1}$    & ${\rm [m\,s^{-2}]}$\\
      $N_x \times N_y \times N_z$ & $64 \times 64 \times 64$ &\\
    \end{tabular}
  \end{ruledtabular}
\end{table}

\begin{figure}[h]
  \centering
  \includegraphics[width=8.5cm]{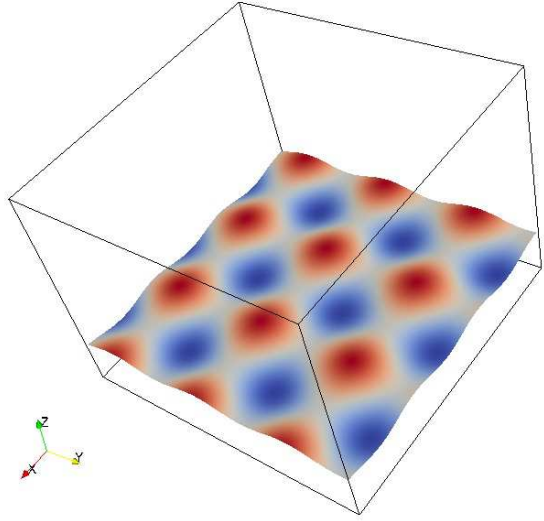}
  \caption{The interface profile for the square-pattern regime at $t=44.5T_v$.
    The interface is colored according to its height:
    the red corresponds to higher region and
    the blue corresponds to lower region.
    Each horizontal length of the domain displayed here is three times that of the calculation
    domain.
    The aspect ratio of this figure is  $(3 L_y) / (3 L_x) = 1.000$.
    Here we show the result with the particle level-set method.}\label{fig:square-pattern}
\end{figure}

\begin{figure}[h]
  \centering
  \includegraphics[width=9cm]{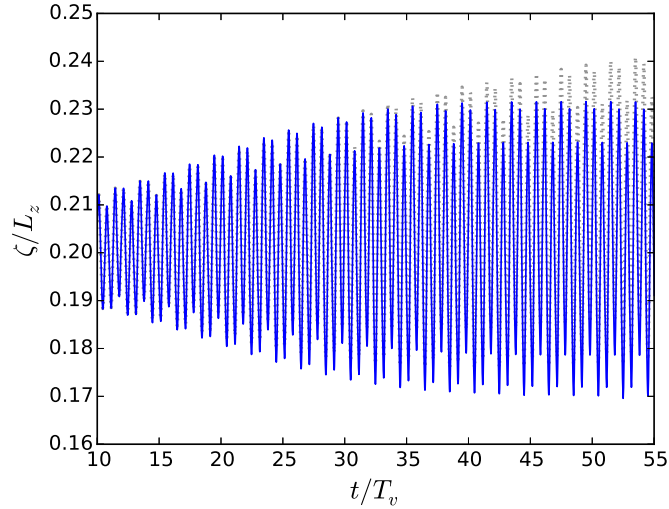}
  \caption{Temporal variation of the interface height at position $(x,~y) =
    (0.5L_x,~0.78L_y)$ for the square pattern calculated with the particle
    level-set method (solid line) and
    with the original level-set method (dotted line). }\label{fig:square-elevation}
\end{figure}

As shown in Fig.~\ref{fig:square-pattern}, indeed a square pattern is
obtained in the simulations.
With both the original level-set method and the particle level-set method,
we start to recognize the square pattern around $t \sim T_v$.
In spite of the same appearance of the pattern, the long-time behaviors
of the two level-set methods are different.
With the particle level-set method, the temporal variation of the interface
elevation at a point $(x,~y) = (0.5L_x,~0.78L_y)$ reaches a steady state
around $t \sim 38 T_v$ as seen in Fig.~\ref{fig:square-elevation} (solid line).
In contrast, with the original level-set method it does not reach a steady state 
but keeps increasing as depicted with the dotted line in
Fig.~\ref{fig:square-elevation}.
However the square pattern is not destroyed
by the unsteadiness up to $t = 45 T_v$, at which we end the simulation.

To characterize the difference between the original and particle level-set methods
we here introduce two time scales:
first pattern recognition time
and
saturation time.
The former is the time we first recognize
the expected pattern, which is $T_v$ for the square pattern case.
The latter is the time needed to reach the steady state, which
is $38 T_v$ for the particle level-set method. Although these time scales
are determined subjectively and are dependent on the initial condition, they play a useful
role in comparison between the two level-set methods as we will discuss later.


Our next target is the hexagonal pattern observed in the experiments
\cite{Arbell2000,Arbell2000a}.
The physical parameters of the simulation are listed in Table~\ref{tab:hexagon-parameter}.
As in the square pattern case, we set the size of the horizontal domain
to the minimal size containing one hexagon.
Specifically,
with the resonant wavevector ${\bm k}_1'$ shown in Fig.~\ref{fig:nonlinear-resonance}(b),
the lengths are $L_x = 2\pi/k_{1x}'$ and $L_y = 2\pi/k_{1y}'$.
The aspect ratio is $L_y/L_x = 0.5774 \approx 1/\sqrt{3}$.
The numbers of grid points used per wavelength are
$(\lambda_1 / L_x) N_x = 0.5 N_x$ and $(\lambda_1 / L_y) N_y = 0.8660
N_y \approx \sqrt{3}/ 2 N_y $.
We use the same initial condition as the square pattern case.

\begin{figure}[h]
  \centering
  \includegraphics[width=8.5cm]{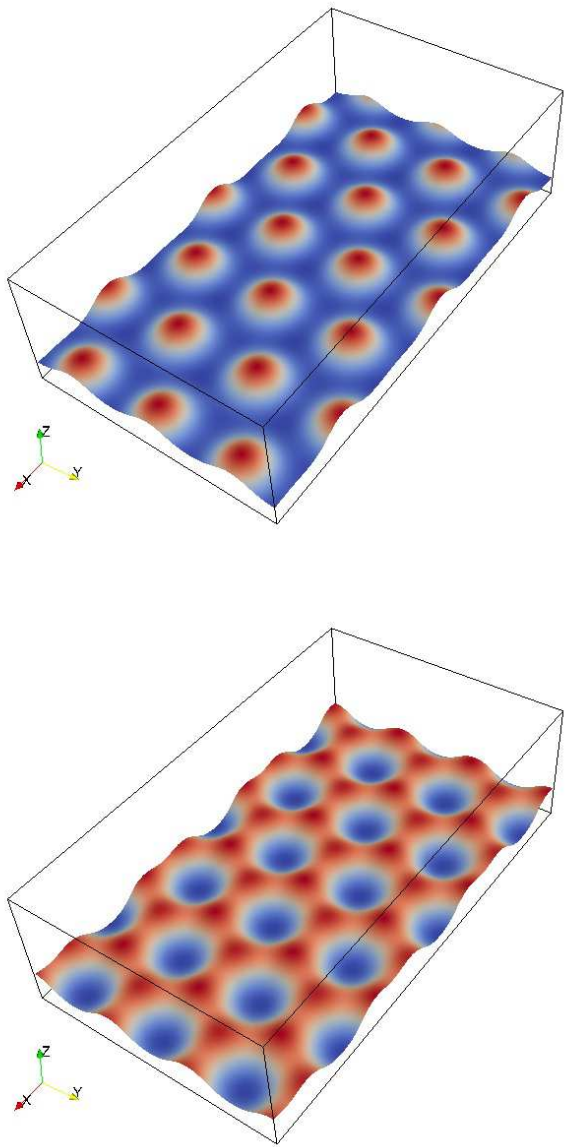}
  \caption{Interface profile for the hexagonal pattern at $t=27.65T_v\text{(above)},
    ~28.22T_v\text{(bellow)}$.
    The interface is colored according to its height:
    the red corresponds to higher region and
    the blue corresponds to lower region.
    The dimension of the domain displayed here is $3 L_x \times 3 L_y \times L_z$.
    The aspect ratio is  $L_y / L_x = 0.5774 \approx 1/\sqrt{3}$.
    Here we show the result with the particle level-set method.
  }
  \label{fig:hexagon-pattern}
\end{figure}

\begin{figure}[h]
  \centering
  \includegraphics[width=9cm]{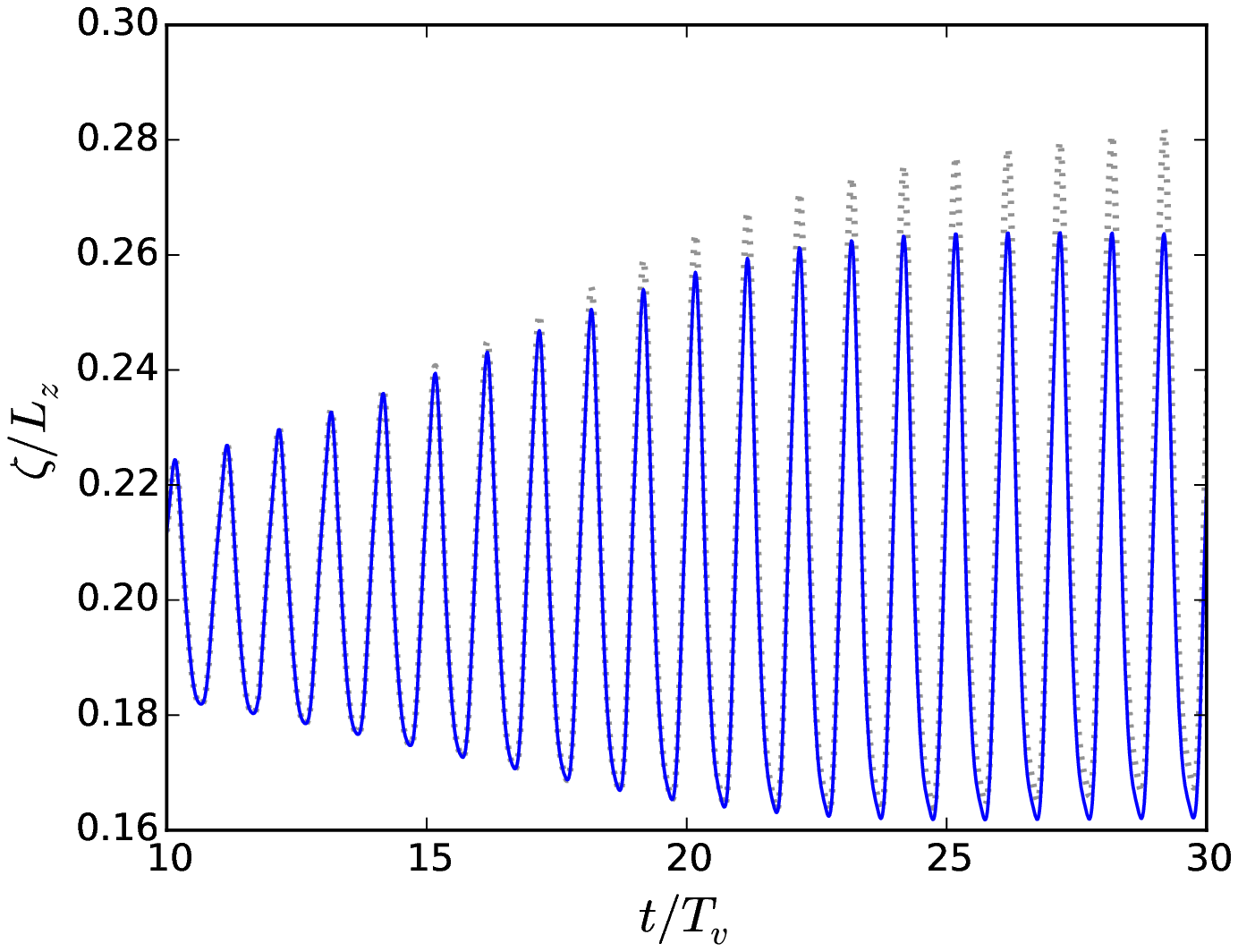}
  \caption{Temporal variation of the interface height at position at $(x,~y) =
    (0.5L_x,~0.5L_y)$ for the hexagonal pattern calculated with the particle level-set method
    (solid line) and with the original level-set method (dotted line).}\label{fig:hexagon-elevation}
\end{figure}

The hexagonal pattern is reproduced with both level-set methods.
The result with the particle level-set method is
shown in Fig.~\ref{fig:hexagon-pattern}.
Despite the pattern being the same, the first pattern recognition time is
different between the two level-set method: $6T_v$ for the original
level-set method and $19T_v$ for the particle level-set method.
The saturation time is $24T_v$ with the particle level-set method
as shown in Fig.~\ref{fig:hexagon-elevation}.
In contrast, saturation does not occur with the original level-set method
during our simulations of length $t = 40T_v$. The hexagonal shape
of the pattern is maintained in spite of the unsteadiness.

The above results on the square and hexagonal patterns suggest
that the original level-set method is not a suitable interface-tracking scheme
for Faraday waves.
Although the patterns initially emerged with the original level-set method
are consistent with the experiment, it is seen that the temporal variation
of the interface height does not reach a steady state.
This unsteadiness in the long run may change the correctly selected pattern initially
into a different shape with the original level-set method.
In the simulation of the rhomboidal pattern, the deficiency of the original
level-set method appears more seriously as we see in the next section.

\begin{table}
  \caption{\label{tab:hexagon-parameter}
    Parameter values for the hexagonal pattern.
    The other parameters are the same as the Table~\ref{tab:linear-stability-parameter}.
  }

  \begin{ruledtabular}
    \begin{tabular}{ccl}
      $L_x$                       & $1.184\times 10^{-2}$     & ${\rm [m]}$     \\
      $L_y$                       & $6.837\times 10^{-3}$    & ${\rm [m]}$     \\
      $A_1$                       & $3.200 \times 10^{1}$     & ${\rm [m\,s^{-2}]}$ \\
      $A_2$                       & $3.000 \times 10^{1}$     & ${\rm [m\,s^{-2}]}$ \\
      $N_x \times N_y \times N_z$ & $112 \times 64 \times 64$ &                 \\
    \end{tabular}
  \end{ruledtabular}
\end{table}

\subsection{Rhomboidal states}
\label{sec:rhom}
\begin{table}
  \caption{\label{tab:rhomboid-parameter}
    Parameter values for the $2k$ rhomboidal states.
    These parameters (except for $L_x, L_y, L_z, \theta, N_x, N_y$ and $N_z$) are identical with
   the experiment by Arbell \textit{et al.}\cite{Arbell2000}
   }
  \begin{ruledtabular}
    \begin{tabular}{ccl}
      $L_x$                       & $1.446\times 10^{-2}$    & ${\rm [m]}$                    \\
      $L_y$                       & $5.234 \times 10^{-3}$   & ${\rm [m]}$                    \\
      $L_z$                       & $1.000\times 10^{-2}$    & ${\rm [m]}$                    \\
      $\rho_t$                    & $1.293$                  & ${\rm [kg \,m^{-3}]}$          \\
      $\rho_b$                    & $9.500\times 10^{2}$     & ${\rm [kg\,m^{-3}]}$           \\
      $\eta_t$                    & $1.822 \times 10^{-5}$   & ${\rm [kg\, m^{-1} \,s^{-1}]}$ \\
      $\eta_b$                    & $2.185 \times 10^{-2}$   & ${\rm [kg\, m^{-1} \,s^{-1}]}$ \\
      $A_1$                       & $2.372 \times 10^{1}$    & ${\rm [m\,s^{-2}]}$            \\
      $A_2$                       & $4.925 \times 10^{1}$    & ${\rm [m\,s^{-2}]}$            \\
      $\omega_1 = 2 \omega_0$     & $3.141 \times 10^{2}$    & ${\rm [s^{-1}]}$               \\
      $\omega_2 = 3 \omega_0$     & $4.712 \times 10^{2}$    & ${\rm [s^{-1}]}$               \\
      $\theta$                    & $0$                      & ${\rm [rad]}$                  \\
      $\sigma $                   & $2.150 \times 10^{-2}$   & ${\rm [kg\,m^{-1}]}$           \\
      $g$                         & $9.807$                  & ${\rm [m\,s^{-2}]}$            \\
      $N_x \times N_y \times N_z$ & $96 \times 32 \times 64$ &                                \\
      bottom-fluid depth          & $2.00 \times 10^{-3} $   & ${\rm [m]}$                    \\
    \end{tabular}
  \end{ruledtabular}
\end{table}

\begin{figure}[h]
  \centering
  \includegraphics{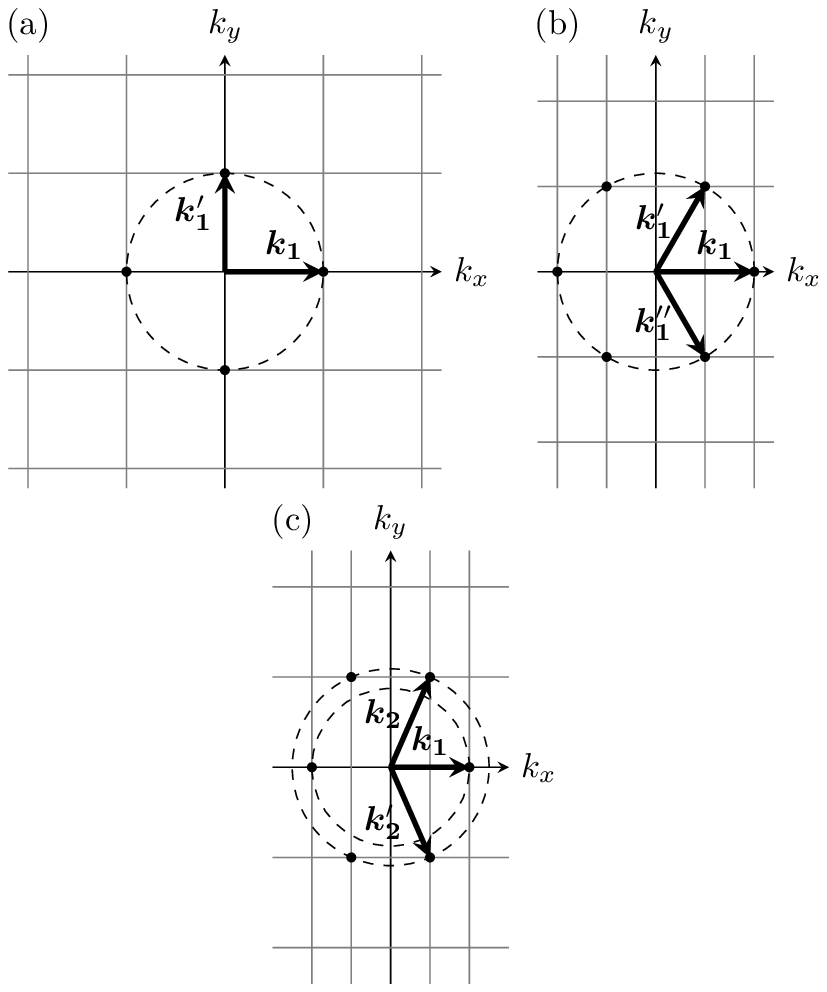}
  \caption{Nonlinear resonant wavevectors
    (a) square pattern: $|\bm{k}_1| = |\bm{k}_1'| = 1.436 \times 10^{3}$ $[m^{-1}]$.
    (b) hexagonal pattern: $|\bm{k}_1| = |\bm{k}_1'| = |\bm{k}_1''| = 1.061
    \times 10^{3}$ $[m^{-1}]$; the angle between $\bm{k}_1'$ and the $k_x$ axis
    is $60^{\circ}$.
    (c) rhomboidal pattern:$|\bm{k}_1|  =8.653 \times 10^{2}$ $[m^{-1}]$;
    $|\bm{k}_2| = |\bm{k}_2'| = 1.275 \times 10^{3}$ $[m^{-1}]$;
    the angle $\varphi$ between $\bm{k}_2$ and the $k_x$ axis
    is $70.16^{\circ}$.
    The grid represents the minimal discretization of the Fourier space
  for each pattern.}
  \label{fig:nonlinear-resonance}
\end{figure}

The next pattern we seek to simulate is called the 2k rhomboidal state
observed in the experiment by Arbell \textit{et al.}\cite{Arbell2000}.
The pattern is observed around the bicritical point which appears as the sharp tip
in Fig.~\ref{fig:linear_analysis}.
There are two linearly
unstable wavenumbers $k_1$ and  $k_2$, hence the name $2k$ rhomboidal states.
As a result of the nonlinear interaction among the resonant modes
a simple resonance relation appears: $\bm{k}_2'
+ \bm{k}_2 = \bm{k}_1$,
as shown in Fig.~\ref{fig:nonlinear-resonance}(c).
This rhomboidal pattern involves two circles in the wavenumber
space, which is a notable difference from the square and hexagonal patterns.

The experiments on the rhomboid patterns
were reported in the two references \cite{Arbell2000,Arbell2000a}.
There is a slight difference in the experimental settings
between the references.
We succeed in simulating the rhomboidal patterns with the same parameters
for each of the two references. However here we present only the result
corresponding to one of the references \cite{Arbell2000}
since it contains a detailed analysis of the pattern
along with a photograph of the rhomboidal pattern.
Note that for the square and hexagonal patterns we use the parameters of
reference~\cite{Arbell2000a}.
The numerical parameters are listed in Table~\ref{tab:rhomboid-parameter}.
We set the size of the horizontal domain again to be minimized containing
one rhomboid, namely $L_x = 2\pi/(k_1/2), L_y = 2\pi/(k_2 \sin\varphi)$ where
$\varphi = 70.16$ is the angle between the vectors ${\bm k}_1$ and ${\bm k}_2$ shown
in Fig.~\ref{fig:nonlinear-resonance}(c).
It is calculated from the relation $2 k_2 \cos\varphi = k_1$.
The aspect ratio is thus $L_y / L_x = 0.3620$.
The numbers of grid points used per wavelength are
$(2 \pi / k_1) (N_x / L_x) = 0.5 N_x$, $ (2 \pi / k_1) (N_y / L_y) = 1.387 N_y $,
$(2 \pi / k_2) (N_x / L_x) = 0.3408 N_x$ and $ (2 \pi / k_2) (N_y / L_y) = 0.9415N_y $.
The initial condition is set in the same way as the cases of the square and
hexagonal patterns.

\begin{figure}[h]
  \centering
  \includegraphics[width=8cm]{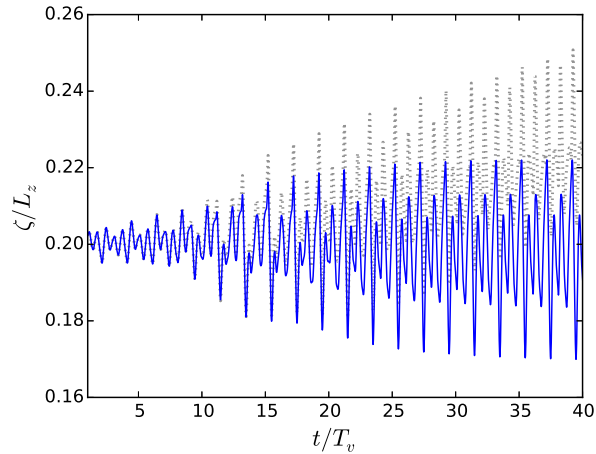}
  \caption{Temporal variation of the interface height at position at $(x,~y) =
    (0.3L_x,~0.25L_y)$ for the rhomboidal state calculated with the particle
    level-set method (solid line) and
    with the original level-set method (dotted line).}
  \label{fig:rhomboid-elevation}
\end{figure}

\begin{figure}[h]
  \centering
  \includegraphics[width=8.5cm]{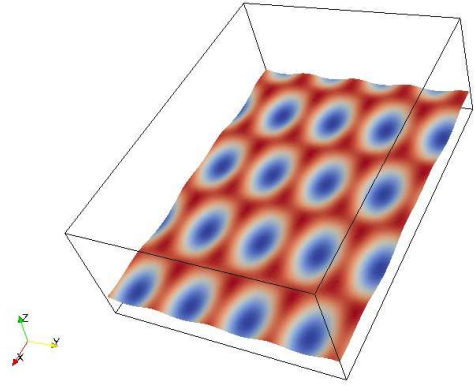}
  \caption{Interface profile for the rhomboidal state at $t=40.38T_v$.
    The interface is colored according to its height:
    the red corresponds to higher region and
    the blue corresponds to lower region.
    The dimension of the domain displayed here is $2 L_x \times 4 L_y \times L_z $.
    The aspect ratio of the displayed domain is  $4 L_y / (2 L_x) = 0.7239$.
    Here we show the result with the particle level-set method.}
  \label{fig:rhomboid-pattern}
\end{figure}

With the original level-set method,
we do not obtain the rhomboidal state.
On the other hand, with the particle level-set method,
we obtain the state as a steady state as shown in
Fig.~\ref{fig:rhomboid-pattern}.
The first pattern recognition time of the rhomboid
with the particle level-set method is $t \sim 29 T_v$ and
the saturation time is the same $t \sim 29 T_v$
as depicted in Fig.~\ref{fig:rhomboid-elevation}.
The first pattern recognition time is much longer than those
of the square and hexagonal patterns.
We consider that the nonlinear interaction among the resonant modes
on the two circles in Fig.~\ref{fig:nonlinear-resonance}(c) takes
a longer time in order to reach a constant oscillation amplitude.

\begin{figure}[h]
  \centering
  \includegraphics[width=8.5cm]{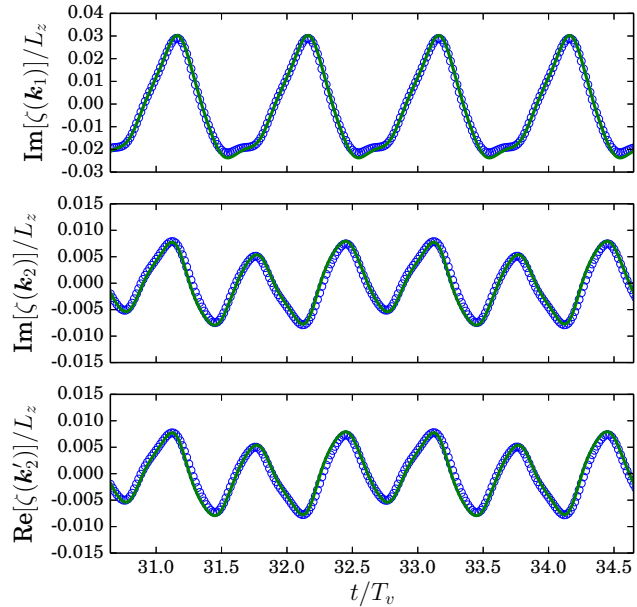}
  \caption{Temporal evolution of the resonance amplitudes of the interface height $\zeta$
  for the wavevectors $\bm{k}_2$, $\bm{k}_2'$, $\bm{k}_1$ (see Fig.~\ref{fig:nonlinear-resonance}(c))
 in the rhomboidal state.
  Here we show only the dominant parts for each wavevector.
  The symbols are data of the simulation with the particle level-set method.
  The solid lines are time evolution of the neutral stable modes calculated with the ten Floquet coefficients in the linear stability analysis.}
\label{fig:temporal-resonance}
\end{figure}

Figure~\ref{fig:temporal-resonance} shows the temporal evolution of the
Fourier amplitudes of the interface height for the three resonant modes.
The circle symbols
represent
the simulation results and
the solid line is the evolution calculated with the Floquet coefficients obtained
in the linear stability analysis \cite{Kumar1994}.
The evolution of the nonlinear rhomboidal modes
(circle symbols in Fig.~\ref{fig:temporal-resonance})
is quite close to that of the linear results, which indicates
that the nonlinear effect in the temporal evolution of the pattern is
weak.

Now we compare the simulation results with
the weakly nonlinear analysis of the rhomboidal
states by Porter \textit{et al.}\cite{Porter2002,Porter2004}.
Their analysis for the first time explains with an elegant broken-symmetry argument
why the rhomboidal pattern appears.
In deriving their amplitude equations up to third order in the amplitude,
they assume
that the rhomboidal state is close to the bicritical point
and
that the damping parameter $\gamma$ is small. Accordingly they expand the
coefficients in the amplitude equations in powers of the vibration
amplitudes $A^*_1 = (A_1-A_{1c}) / A_{1c}, A^*_2 = (A_2-A_{2c}) / A_{2c}$
and the damping parameter $\gamma / \omega_0$.
The resulting coefficients of the quadratic term of the amplitude, their signs
and dependence on $\gamma$, explain the rhomboidal pattern selection for certain
frequency ratios $m:n$.
However, as they discussed,
it is not clear that the damping parameter $\gamma / \omega_0$
is small enough in the experiments \cite{Arbell2000}.

\begin{figure}[h]
  \centering
  \includegraphics[width=8.5cm]{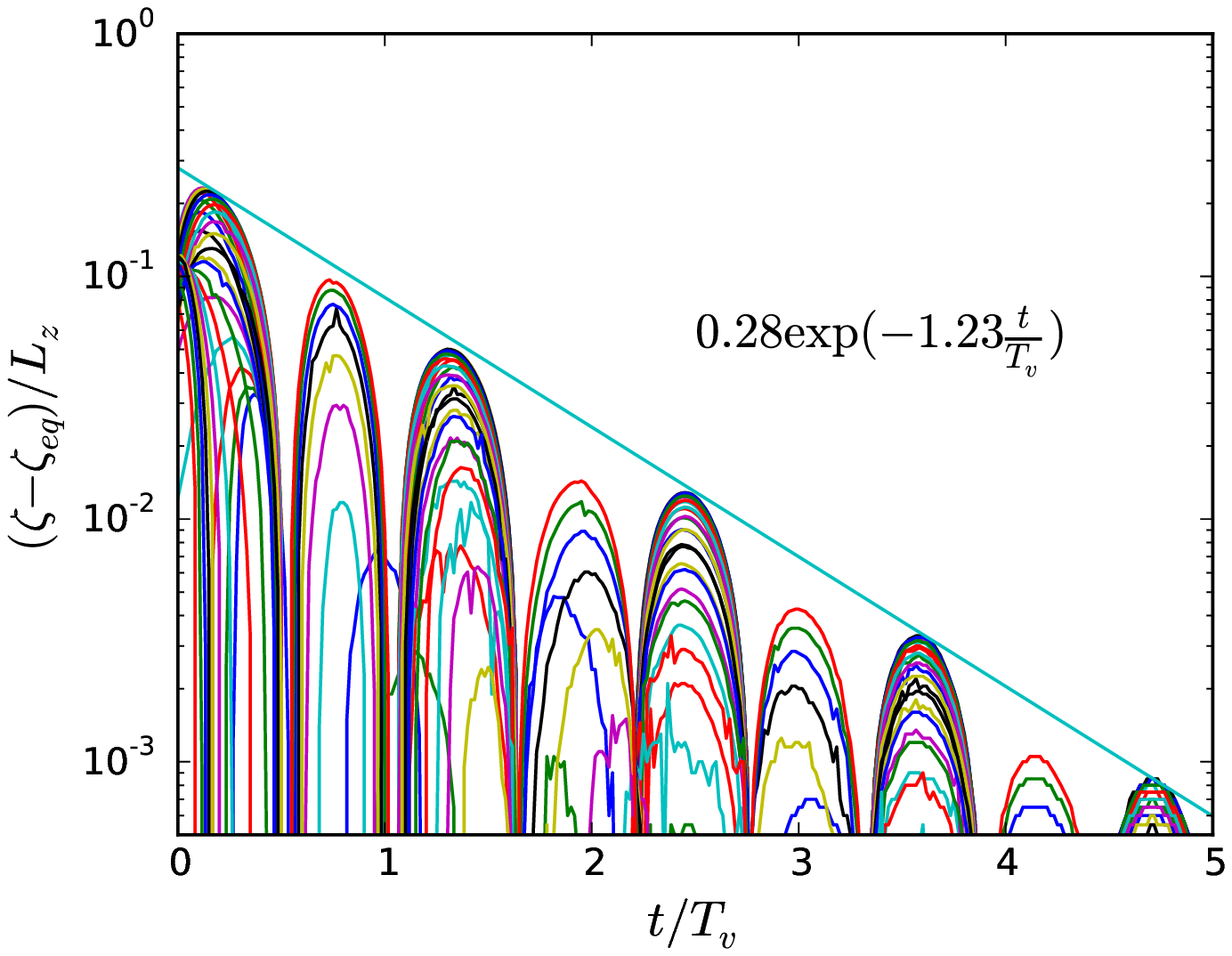}
  \caption{Measurement of the damping parameter $\gamma$. The curves are
 temporal variations of the interface height at various points on the line $x = 0$ without
 the vibration forcing. The damping parameter $\gamma / \omega_0$ can be estimated from
 the envelope $0.28\exp(-1.23t / T_v)$. The interface height at the quiescent state
 (where the Faraday waves decay completely) is denoted as $\zeta_{eq}/L_z \simeq 0.197$, which is slightly different from the initial
 interface height $0.20L_z$ without the random perturbation.}
  \label{fig:gamma}
\end{figure}

To test the assumption, we measure the damping parameter from our simulation
data. Before doing this we calculate it with dimensional analysis:
the damping parameter of the bottom fluid
can be estimated as $\gamma(k) / \omega_0 = 2 \eta_b / (\rho_b k^2 \omega_0)$
with the critical wavenumber $k$.
This gives $\gamma(k_1) / \omega_0 = 0.219$ and $\gamma(k_2) / \omega_0 = 0.476$,
where the critical wavenumbers $k_1$ and $k_2$ are determined for
the critical vibration amplitudes  $A_{1c} = 21.9$ and $A_{2c} = 44.8$.
These dimensional values can differ in orders of magnitudes
from the actual damping parameter.
In our nonlinear simulation of the rhomboidal pattern,
we set the normalized vibration amplitudes
$A^*_1 = (23.7 - 21.9) / 21.9 = 0.0831$ and
$A^*_2 = (49.3 - 44.8) / 44.8 = 0.100$, which justifies
the expansion in terms of $A^*_1$ and $A^*_2$ in the coefficients of
the weakly nonlinear analysis.
In order to estimate the damping parameter in our nonlinear simulation,
we used a method similar to that used in the experiment \cite{Cocciaro1991}:
we take the snapshot at $t = 141.4 T_v$ from the rhomboidal pattern simulation
as an initial condition;
we then start the simulation without the vibration forcing
and measure how the interface elevation decays in time.
The temporal interface behaviors on the
line at $x/L_x=0$ are shown in Fig.~\ref{fig:gamma}.
The envelope in the figure gives $\gamma T_v \sim 1.23$.
Consequently, the damping parameter $\gamma /\omega_0$ is $0.196$.
Although this is smaller than unity,
it may not be small enough to ignore its higher order terms.

We next try to obtain the slowly varying amplitudes from
the fully nonlinear evolution of the resonant modes shown in Fig.~\ref{fig:temporal-resonance}.
For example, Im$\zeta({\bm k}_2)$  divided by the sub-harmonic oscillation
$C \sin[2\pi/(\omega_2/2)~ t + \Theta]$,
where $\omega_2 = 3\omega_0$, $C$ is a suitable factor and $\Theta$ is a suitable phase,
should give a slowly evolving function.
We divided the resonant mode (symbols in Fig.~\ref{fig:temporal-resonance})
by the sub-harmonic oscillation.
However the calculated function is not slowly varying in time.
Moreover, we divided the nonlinear data (symbols)
by the linear Floquet-mode data (lines) in Fig.~\ref{fig:temporal-resonance}.
The calculated function is not slowly varying either.
Nevertheless we look at the phase-space orbit formed by the three variables
in Fig.~\ref{fig:temporal-resonance}. We do not find a characteristic structure
often associated with the solutions of the normal-form equations corresponding
to the rhomboidal structure.
Hence we are not able to compare our data with the weakly nonlinear analysis
in this respect.

\subsection{Comparison between original and particle level-set
  method}\label{SEC:comp-with-orig}

We observe that
the original level-set method and
the particle level-set method yield
qualitatively different results.
With the original level-set method,
the square and hexagonal patterns are observed
but do not become constant-amplitude oscillations.
The rhomboidal state, which is here the main target,
is not observed.
On the other hand, in our simulation with the particle level-set method,
all three patterns
are observed and become constant-amplitude oscillations in agreement with
the experiments.
This difference is due to the well-known problem of the original
level-set method, which we discuss here.

\begin{figure}
  \centering
  \includegraphics[width=8cm]{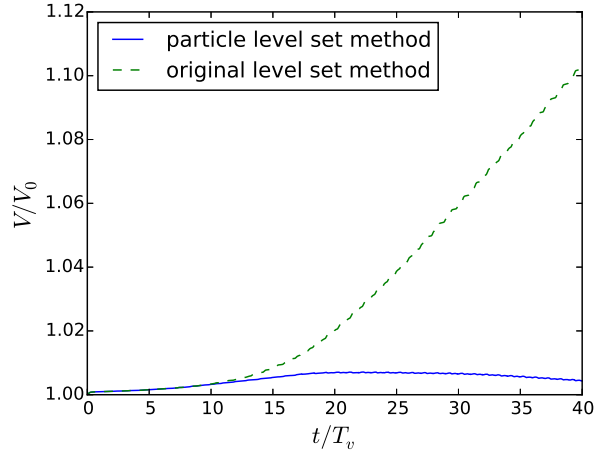}
  \caption{Comparison of variation of the bottom fluid volume between the
 original level-set  method and the particle level-set method in the case of the hexagonal pattern.}
  \label{fig:hexagon-volume}
\end{figure}

\begin{figure}[h]
  \centering
  \includegraphics[width=8cm]{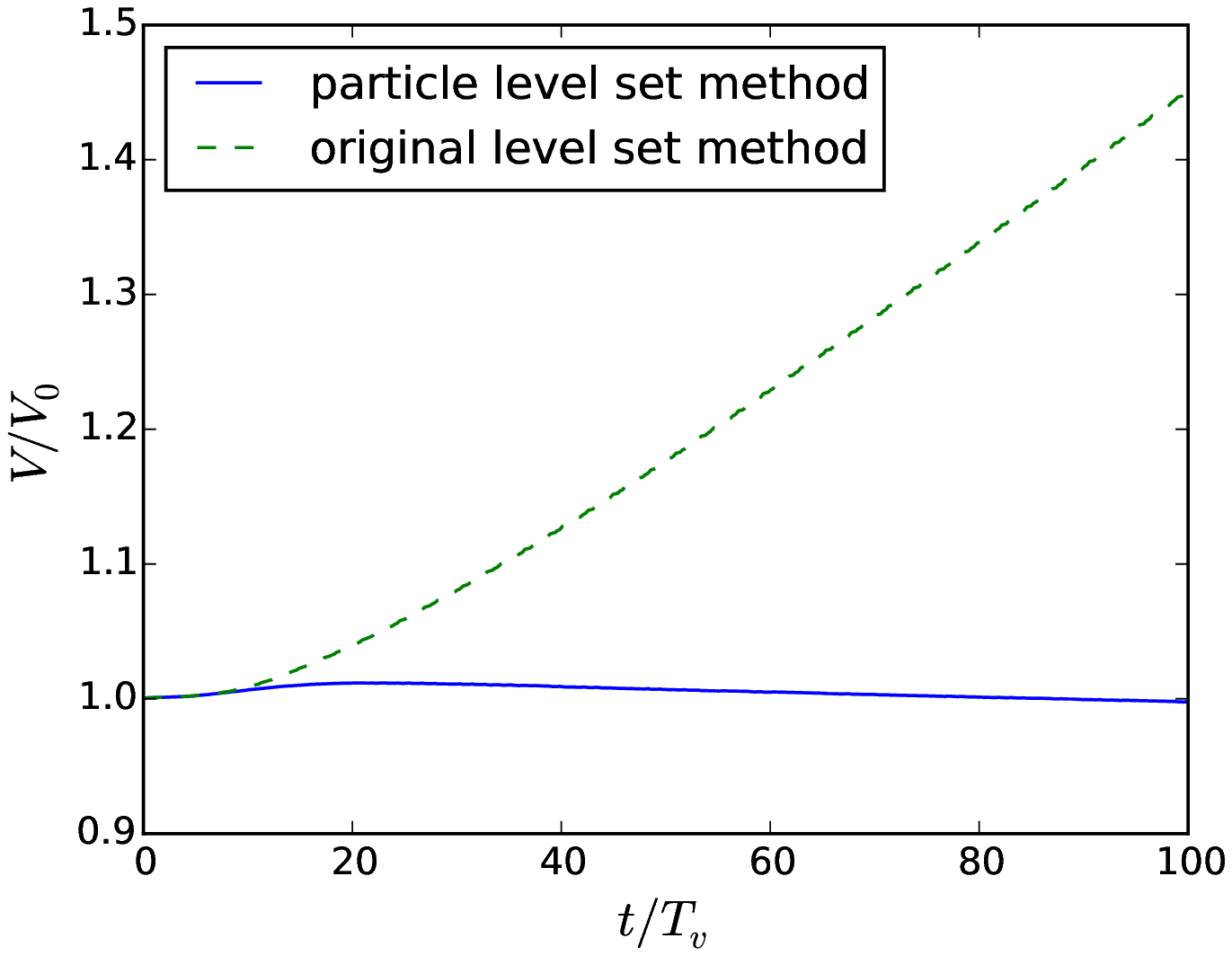}
  \caption{Same as Fig.~\ref{fig:hexagon-volume} but for the case of the rhomboidal states.}
  \label{fig:rhomboid-volume}
\end{figure}

In order to clarify the difference between the two level-set methods,
we look at how well the volume of the lower fluid is conserved during the time evolution.
The volume of lower fluid is
calculated with $H(\phi)$, Eq.~(\ref{eq:2}), as $V(t) = \int H(\phi(\bm{x}, t)) d{\bm x}$.
The variations of the volume for the hexagonal and rhomboidal cases
are shown in Figs.~\ref{fig:hexagon-volume}, \ref{fig:rhomboid-volume}.
with the numerical parameters listed in Tables~\ref{tab:hexagon-parameter}, \ref{tab:rhomboid-parameter}.

As shown in Figs.~\ref{fig:hexagon-volume} and \ref{fig:rhomboid-volume},
the volume increases with the original level-set method, instead of being
conserved. This non-conserving property of the original level-set method is well
known\cite{Andrea2009,Enright2002,Sussman2000,Sussman1999a}.
This explains why the interface height does not reach constant-amplitude
oscillations with the original level-set method for the square and hexagonal
patterns.
Concerning the rhomboidal pattern, the original level-set method fails
to exhibit the pattern. But with the particle level-set we start
to recognize rhomboids  at $t = 29 T_v$ (first recognition time).
At this time it is seen
from Fig.~\ref{fig:rhomboid-volume} that the volume in the simulation with
the original level-set method increases by 10\%.
In other words, a long time is needed
for the nonlinear interaction to form the resonant modes for the rhomboidal
pattern. During this time, the error of the simulation with the original level-set method,
the increase of the bottom-fluid volume, becomes so significant that the rhomboidal
pattern is not observed.
Therefore we conclude that the particle level-set method is more suitable
than the original level-set method
to reproduce complex patterns such as the rhomboidal pattern,
which require a long time for selection

\section{Summary and discussion}\label{sec:concluding-remarks}
Motivated by the recent experiments of Faraday waves with two or more
frequency forcings exhibiting even richer patterns than the single frequency case,
we have conducted a numerical simulation of the two-frequency Faraday waves,
specifically targeting the rhomboidal pattern.

We first validated our numerical simulation with the linear stability analysis of
the two-frequency Faraday waves \cite{Besson1996}.
The two simple patterns, the square and hexagonal patterns, in the nonlinear regime
were simulated with the same physical parameters as the experiment \cite{Arbell2000a}.
In particular, the simulation using the particle level-set method in the
minimal computational domain reproduced the two patterns in agreement with the experiment.
Employing the particle level-set method, we finally reproduced numerically
the $2k$ rhomboidal states, the most complex pattern in this numerical study,
with fluid properties identical to those of the experiments.
We next checked whether the rhomboid obtained in our simulation satisfies
the assumption made in the weakly nonlinear analysis for the rhomboidal
pattern\cite{Porter2002, Porter2004}. Specifically, the assumption concerns
the smallness of the damping parameter and the vibration amplitudes.
We found that the damping parameter of the rhomboidal pattern in our simulation
is marginally small. Further comparison with the weakly nonlinear analysis is
difficult.

In these simulations, we used two level-set methods: the original level-set
method and the particle level-set method.
The interface motion of the Faraday waves appears quite
modest in the sense that it is not usually considered as a typical target
of the interface-tracking schemes.
One may think that any modern scheme is capable of simulating Faraday waves.
However, due to the well-known problem of the original level-set method\cite{Andrea2009},
we failed to simulate the square and hexagonal patterns as steady states
and to reproduce the rhomboidal pattern at all.
Thus the Faraday wave problem requires an accurate scheme tracking
of the interface such as the particle level-set method. One reason
for this is that we need to simulate the system for a long time if
we start with a random initial condition.
We believe that, in developing a new implementation of the interface-tracking scheme,
the Faraday wave problem can be a benchmark problem in addition to a physical phenomenon.
In the linear regime quantitative comparison can be made as demonstrated in
the simulation by P\'{e}rinet \textit{et al.}\cite{Perinet2009}.
In the nonlinear regime qualitative comparison can be made (whether or not the
right pattern emerges if we choose parameters for a certain pattern observed in experiments).

We carried out  simulations on the minimal calculation domains
to reproduce the three patterns. Its effect was studied for the rhomboidal case
in the following way.
Simulations were run in domains which were twice as large with twice as many
points, thus keeping the density of numerical grid points constant.
Accordingly we have the same grid spacings in the physical
space, $L_x/N_x$ and $L_y/N_y$.
(for the $z$ direction we keep the same values for $L_z$ and $N_z$).
The rhomboidal pattern is observed with this setting with
the same first pattern recognition time and the saturation time.
Hence it is unlikely that the minimal domain setting affects
the pattern selection numerically.
We also checked whether the number of grid points in the vertical direction
$N_z$ is sufficient or not by doubling $N_z$ but retaining the other
parameters as in Table~\ref{tab:rhomboid-parameter}. The result does not
change.

As we mentioned briefly in the Introduction, many other
patterns are observed in the experiments of the two-frequency forced Faraday waves.
In fact, our initial goal was to reproduce not only the rhomboidal
pattern but also the hexagonal based oscillon (HBO)
(also known as double hexagonal superlattice (DHS)),
the spatially subharmonic superlattice
pattern (SSS) and the oscillon observed experimentally by Arbell \textit{et al.}\cite{Arbell2000,Arbell2000a}.
So far, we have not been able to reproduce those patterns perhaps due to our strategy to
use the minimal calculation domain including one pattern.
Setting the minimal domain
corresponds in terms of the Fourier space to
maximizing the grid spacings in the $k_x$ and $k_y$ directions
to include the resonant modes with discretized points.
For these patterns we failed to simulate; in fact,
it is not clear how to set a minimal domain even with the knowledge of
the selected resonant modes available from the experiments.

Now we take the HBO pattern as an example and discuss the difficulty of
setting the minimal domain.
In the linear analysis of the HBO case, two different wavenumbers simultaneously
become unstable.
Hence, as in the case of the rhomboidal pattern shown in Fig.~\ref{fig:nonlinear-resonance},
two circles can be important.
However according to the experiment the resonant modes lie on only one of the
two, which we call the resonant circle; we call the other circle the non-resonant circle.
As a first trial we took the minimal calculation domain to resolve only these resonant modes
on the resonant circle without including modes on the non-resonant circle.
With this minimal domain and the particle level-set method,
we did not obtain the HBO pattern at all starting from the same initial condition as in Sec.~\ref{sec:numerical-results}.
We speculate that, in the course of establishing the resonant modes,
the modes on the non-resonant circle are important in the pattern selection
and hence should be taken into account properly in the simulation.
Of course, if we could enlarge the calculation domain and increase
the number of grid points in the physical space in order to take
a large number of mesh points near the non-resonant circle in the Fourier
space, this problem might be overcome.
Even though we have doubled $L_x$, $L_y$, $N_x$ and $N_y$,
in order to take smaller grid spacing
in the Fourier space, the HBO pattern  
did not emerge.
A finer grid would make the cost of computation prohibitively high
(notice that a long time simulation is also needed here).

We also ran the simulation of the SSS but failed possibly
for the same reason.
The resonant modes of the SSS observed experimentally lie either
on a circle whose wavenumber (radius) is linearly stable
or one of the two circles determined by the linear stability analysis.
For both the HBO and the SSS cases, we checked that the volume is conserved
to the same degree as it is in the rhomboidal case with the particle level-set method.
Regarding the oscillon, the structure of the resonant modes in the Fourier space
is not clarified experimentally, implying that we do not have any guidance on the
discretization of the Fourier space.
Perhaps, guessing from the physical-space  appearance
of the oscillon, the number of excited Fourier modes is
very large compared with other patterns.
To circumvent this sort of difficulty, a completely different numerical scheme
with Chebychev polynomials
for capturing a localized structure is proposed by Lloyd {\textit et
 al.}\cite{Lloyd2005}, which  may be worth exploring.
Moreover the oscillon's metastability \cite{Arbell2000a} may make simulation even more challenging.
Our future work is an approach relying on computing power in which we take as high a resolution
as possible to reproduce complex patterns like
the HBO, the SSS and the oscillon.
We believe that, if such a simulation succeeds, it would provide
knowledge about the role of the modes on the non-resonant circles in pattern
selection.

\begin{acknowledgments}
  This work is supported by the grant for JSPS fellows No. 25$\cdot$1056
  and by the JSPS KAKENHI (C) No. 25400400.
  We are grateful to Professor Sadayoshi Toh for his continuous encouragement.
  We thank anonymous referees for comments and for drawing our attention to
  the two references~\cite{Wagner2003,Lloyd2005}.
\end{acknowledgments}

\bibliography{extracted}



%
%

%



\end{document}